\documentclass[showpacs,prl,twocolumn,aps]{revtex4}

 \usepackage{bm}
 \usepackage{amsmath}
 \usepackage{amssymb}
 \usepackage{latexsym} 
 \usepackage{amsfonts} 
 \usepackage{epsfig}
 \usepackage{psfrag} 

 \newcommand{\ket}[1]{\left|#1\right>}

 \newcommand{\ul}{\underline} 
 \newcommand{\f}[1]{\mbox{\boldmath$#1$}}

 \newcommand{\na}{\mbox{\boldmath$\nabla$}}
 \newcommand{\bea}{\begin{eqnarray}}
 \newcommand{\ea}{\end{eqnarray}}
 \newcommand{\eea}{\end{eqnarray}}
 \newcommand{\ord}{{\cal O}}

\begin{document}

\title{Dynamical zero-temperature phase transitions and 
cosmic inflation/deflation} 

\author{Ralf Sch\"utzhold}
\affiliation{Institut f\"ur Theoretische Physik, 
Technische Universit\"at Dresden, D-01062 Dresden, Germany}

\begin{abstract} 
For a rather general class of scenarios, sweeping through a
zero-temperature phase transition by means of a time-dependent
external parameter entails universal behavior:
In the vicinity of the critical point, excitations behave as 
quantum fields in an expanding or contracting universe. 
The resulting effects such as the amplification or suppression of quantum
fluctuations (due to horizon crossing, freezing, and squeezing)  including 
the induced spectrum can be derived using the curved space-time analogy. 
The observed similarity entices the question of whether cosmic inflation 
itself might perhaps have been such a phase transition.
\end{abstract} 


\maketitle

\begin{figure}[ht]
\centerline{\mbox{\epsfxsize=5.5cm\epsffile{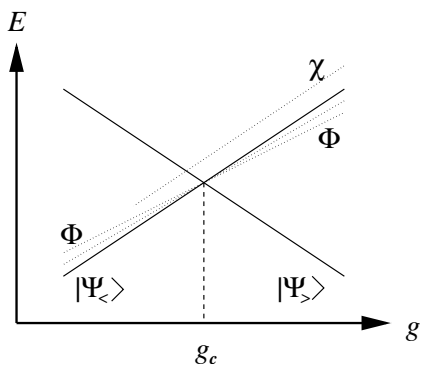}}}
\caption{Sketch of level structure near the critical point.\\
The energy $E$ of the levels is plotted as a function of some 
external parameter $g$. 
At the critical point ${g=g_c}$, the ground state changes 
from $\ket{\Psi_<}$ to $\ket{\Psi_>}$. 
Typical excitations above the ground state $\ket{\Psi_<}$
for ${g<g_c}$ are denoted by $\Phi$ and $\chi$.
Some of the excitations $\chi$ may remain stable after crossing 
the critical point $g=g_c$, whereas others $\Phi$ become unstable 
i.e., lie below $\ket{\Psi_<}$.}
\label{picture}
\end{figure}

In contrast to thermal phase transitions occurring when the strength
of the thermal fluctuations equals a certain threshold 
(and so changes the character of the stable phase), zero-temperature 
phase transitions such as quantum phase transitions~\cite{sachdev}
denote the crossover of different ground states at a certain
(critical) value of some external parameter, where quantum
fluctuations play a dominant role, cf.~Fig.~\ref{picture}. 
In both cases, there exists a vast amount of literature regarding the 
equilibrium properties in the vicinity of the phase transition, for
example in view of universal behavior (e.g., scaling laws) near the
critical point.
However, since response times typically diverge in the vicinity of
the critical point, sweeping through the phase transition with a finite
velocity, for example, leads to a break-down of adiabaticity and thus 
might generate interesting dynamical (non-equilibrium) effects. 
For thermal phase transitions, a prominent example is the Kibble-Zurek
mechanism, i.e., the generation of topological defects via rapid
cooling (quench), which can be applied to the phase transitions in the
early universe as well as in the laboratory~\cite{kibble}. 

The following considerations are devoted to non-equilibrium effects in
a rather general class of zero-temperature transitions, which also
expose a remarkable analogy between cosmology and laboratory physics, 
see also~\cite{zurek}.
Let us consider a quantum system (at zero temperature) described by
the Hamiltonian $\hat H$ depending on some external parameter $g$.
At a certain critical value of this parameter $g_c$, the system is
supposed to undergo a phase transition, i.e., the ground 
state $\ket{\Psi_<(g)}$ of $\hat H(g)$ for $g<g_c$ is different from
the ground state $\ket{\Psi_>(g)}$ of $\hat H(g)$ for $g>g_c$. 
For example, $\ket{\Psi_<(g)}$ and $\ket{\Psi_>(g)}$ could have
different global/topological properties (such as magnetization) 
in the thermodynamic limit.  

Before the phase transition $g<g_c$, the phase space can be 
explored in terms of quasi-particle excitations above the ground 
state $\ket{\Psi_<(g)}$. 
If we make the assumption that the Hamiltonian $\hat H$ is analytic in
the external parameter $g$, we may follow the behavior of these
excitations through the critical point $g_c$ and extrapolate them to
the region $g>g_c$.  
For $g>g_c$, the state $\ket{\Psi_<(g)}$ is no longer the ground state
and hence some of the excitations must become unstable after the 
critical point $g>g_c$, since now $\ket{\Psi_>(g)}$ has a lower energy
than $\ket{\Psi_<(g)}$, cf.~Fig.~\ref{picture}.
In the following, we shall focus on these quasi-particle excitation
modes, whose energy is positive for $g<g_c$ and becomes negative after
crossing the critical point $g>g_c$.
In order to describe these excitations quantitatively, a few additional
assumptions/approximations are necessary:
\\
{\bf Linearity:} 
We assume that the crossover from stability to instability of these
modes can be described using linear stability analysis, i.e., that the 
associated quantum fluctuations are small enough.
Consequently, the onset of instability occurs if the dispersion
relation $\omega^2(k)$ dives below the $k$-axis. 
\\
{\bf Vanishing gap:} 
Since we are mainly interested in low-energy and long-wavelength 
excitations (which will turn out to yield universal behavior 
largely independent of the microscopic structure), we shall assume 
that the modes $\Phi$ are gap-less such as Goldstone modes, i.e., 
${\omega^2(k=0)=0}$ for all $g$. 
The analogy to quantum fields in an expanding/contracting universe 
applies to the general case with a gap as well, but the main points 
of interest (analogue of cosmic horizon) can be studied for gap-less 
modes.
\\
{\bf Analyticity:} 
The dispersion relation $\omega^2(k,g)$ is supposed to be an analytic
function of $k$.  
Hence a Taylor expansion starts with the term 
[with ${c^2(g<g_c)>0}$]
\bea
\label{dispersion}
\omega^2(k,g)=k^2c^2(g)+\ord(k^3)
\,.
\ea
{\bf Independence:} 
As the final ingredient, we assume that the unstable modes are
independent of each other such that it suffices to consider one scalar
mode $\Phi$. 

Based on these assumptions, we may construct the low-energy effective
theory describing the mode $\Phi$ in the vicinity of the critical
point $g_c$ according to the dispersion relation~(\ref{dispersion}).
As demonstrated in Ref.~\cite{visser}, given the above conditions, 
the most general low-energy effective action can be cast into the form 
\bea
{\cal L}_{\rm eff}=\frac{1}{2}\,\sqrt{-g_{\rm eff}}\,
g_{\rm eff}^{\mu\nu}(\partial_\mu\Phi)(\partial_\nu\Phi)
\,,
\ea
where $\sqrt{-g_{\rm eff}}\,g_{\rm eff}^{\mu\nu}$ denotes a matrix 
depending on the system under consideration as well as the external 
parameter and a sum convention over ${\mu,\nu=0\dots3}$ is employed. 

Hence the excitations $\Phi$ behave as minimally coupled and mass-less 
scalar quantum fields in curved space-times whose geometry is determined 
by the effective metric $g_{\rm eff}^{\mu\nu}$; 
which is basically the underlying idea of the analogue gravity
concept, see, e.g., \cite{unruh,droplet,artificial}. 
If we assume the quantum system under consideration to be effectively 
homogeneous and isotropic at large wavelengths $\lambda$, 
the above action simplifies to
\bea
\label{action}
{\cal L}_{\rm eff}=\frac{1}{2}
\left(\frac{1}{\alpha}\,\dot\Phi^2-\beta\,(\na\Phi)^2\right)
\,.
\ea
The external parameters $\alpha$ and $\beta$ may depend on $g$ and hence 
on time and must be non-negative before the phase transition. 
The convenience of the choice of $1/\alpha$ in the first term becomes
apparent after constructing the associated effective Hamiltonian 
\bea
{\cal H}_{\rm eff}=\frac{1}{2}
\left(\alpha\,\Pi^2+\beta\,(\na\Phi)^2\right)
\,.
\ea
In the homogeneous and isotropic case, the effective metric reads 
(for ${g<g_c}$)
\bea
\label{ds}
ds^2_{\rm eff}=
\sqrt{\alpha\beta^3}\,dt^2-\sqrt{\beta/\alpha}\,d\f{r}^2
\,.
\ea
Since the mode $\Phi$ becomes unstable for $g>g_c$, at least one of
the two parameters has to change its sign at the critical point. 
There are basically three possibilities:
\\
{\bf A:} $\alpha\downarrow0$ while $\beta$ remains finite.
(Note that $1/\alpha\downarrow0$ is not possible since the propagation 
speed would diverge.)
According to Eq.~(\ref{ds}), the effective metric corresponds to an 
expanding universe in this case. 
\\
{\bf B:} $\beta\downarrow0$ while $\alpha$ remains finite
(similarly, $1/\beta\downarrow0$ is not possible),
which corresponds to a contracting universe.
\\
{\bf C:} Both $\alpha$ and $\beta$ become singular.

Having established the analogy to cosmology, we may now apply the
tools and concepts known from curved space-times~\cite{birrell}.
The space-time described by the above metric contains a (particle)
horizon if the maximum distance (measured in co-moving coordinates) 
which can be traveled starting at the time $t_{\rm in}$
\bea
\Delta r
=
\int\limits_{t_{\rm in}}^{t_{\rm out}} dt\,c(t)
=
\int\limits_{t_{\rm in}}^{t_{\rm out}} dt\,\sqrt{\alpha(t)\beta(t)}
\,,
\ea
is finite, i.e., points beyond the horizon size $\Delta r$ can never
be reached. 
Obviously, since $c=\sqrt{\alpha(t)\beta(t)}$ is bounded from above, 
a horizon always exists if the critical point is reached in a
finite laboratory time $t_{\rm out}$
(which might still correspond to an infinite proper time in cosmology), 
but also if $\alpha$ and/or $\beta$ decrease fast enough 
for $t_{\rm out}\uparrow\infty$, for example $\alpha\propto1/t^n$ 
with $n>2$, cf.~\cite{2compBEC}.
 
As one can infer from the above equation, the horizon size 
(in terms of the laboratory coordinates $t,\f{r}$) always decreases 
as $t_{\rm in}$ increases.
Hence, all $\Phi$-modes with a given (effective) wavelength $\lambda$ 
cross the horizon at some time 
(when $\lambda$ exceeds the horizon size $\Delta r$).
After this horizon crossing, the modes cannot oscillate anymore 
(freezing) due to loss of causality across the horizon and their
quantum state gets squeezed.
This mechanism is completely analogous to the amplification of quantum
vacuum fluctuations of the inflaton field in our present standard
model of cosmology -- which are supposed to be the seeds for structure
formation. 

In order cast this analogy into a more quantitative form, let us consider 
the equations of motion for $\Phi$
\bea
\label{eom-phi}
\left(
\frac{\partial}{\partial t}\,\frac{1}{\alpha(t)}\,\frac{\partial}{\partial t}
-\beta(t)\,\na^2
\right)\Phi=0
\,.
\ea
In the homogeneous and isotropic case $\na\alpha=\na\beta=0$, there is
a duality between the field $\Phi$ and its canonical 
momentum ${\Pi=\dot\Phi/\alpha}$ since $\Pi$ obeys the same the
equation of motion as $\Phi$ with $\alpha$ and $\beta$ interchanged 
\bea
\label{eom-pi}
\left(
\frac{\partial}{\partial t}\,\frac{1}{\beta(t)}\,\frac{\partial}{\partial t}
-\alpha(t)\,\na^2
\right)\Pi=0
\,.
\ea
This duality is analogous to that between the electric and magnetic
field in the absence of macroscopic sources. 
However, even though the equations of motion in cases {\bf A} and {\bf B} 
are related by this duality, the behavior of the quantum fluctuations 
(e.g., their spectrum) is different.

Assuming general power-law time-dependence near the critical 
point $\alpha\propto t^a$ and $\beta\propto t^b$, we may remember 
the coordinate invariance of the effective geometry and introduce 
another time coordinate $\tau$ proportional to $t^{2/(2+a+b)}$, 
for which the wave equation simplifies to 
\bea
\left(
\frac{\partial^2}{\partial\tau^2}
-\frac{2\nu-1}{\tau}\,\frac{\partial}{\partial\tau}
-\na^2\right)\Phi=0
\,,
\ea
provided that $\nu$ is given by
\bea
\label{nu}
\nu=\frac{1+a}{2+a+b}
\,.
\ea
Hence the solutions for the spatial Fourier modes can be expressed 
in terms of Hankel functions~\cite{a+s}
\bea
\label{hankel}
\Phi_k^\pm(\tau)=\tau^\nu\,H^\pm_\nu(k\tau)
\,.
\ea
Sufficiently far away from the critical point $|\tau|\uparrow\infty$, 
the modes 
oscillate ${\Phi_k^\pm(\tau)\sim\tau^\nu\,e^{\pm ik\tau}/\sqrt{k\tau}}$
and the solutions $\Phi_k^\pm$ are basically the functions in the 
expansion into creation and annihilation operators corresponding to 
the adiabatic vacuum state, cf.~\cite{birrell}.
When approaching the phase transition, however, adiabaticity 
breaks down and the modes cross the horizon and freeze~\cite{a+s}
\bea
\Phi_k^\pm(|\tau|\downarrow0)\sim k^{-\nu}
\,.
\ea
I.e., the spectrum of the two-point correlation 
function ${\langle\hat\Phi(\f{r})\hat\Phi(\f{r'})\rangle}$ of the
frozen quantum fluctuations is determined by the parameter $\nu>0$ in 
Eq.~(\ref{nu}).

Let us examine a few examples:
The trivial case ${a=b=0}$ of course reproduces the undisturbed 
spectrum ${\nu=1/2}$.
Neglecting the back-reaction of the quantum fluctuations onto the 
the dynamics of the external parameter $g$, its time-evolution $g(t)$ 
should not experience anything special at the critical 
point ${g=g_c}$ and hence the most natural choice for its
dynamics is a constant velocity $g-g_c \propto t$. 
If we assume the effective Hamiltonian to be an analytic function 
of $g$, this implies $a=1$ and/or $b=1$.  
(Otherwise $a$ and $b$ are the characteristic exponents $|g-g_c|^a$ 
and $|g-g_c|^b$ occurring in $\hat H$.)
In case~{\bf A} (expanding universe) with $a=1$ and $b=0$, 
we obtain ${\nu=2/3}$, i.e., quantum fluctuations with large 
wavelengths are amplified. 
case~{\bf B} (contracting universe) with $a=0$ and $b=1$ 
yields ${\nu=1/3}$, i.e., quantum fluctuations with large wavelengths 
are suppressed.
Finally, ${a=b=1}$ (case~{\bf C}) leads to an undistorted 
spectrum ${\nu=1/2}$, which is not surprising once one realizes that
the effective metric in Eq.~(\ref{ds}) is exactly flat in terms of the 
new time-coordinate ${ds^2_{\rm eff}=d\tau^2-d\f{r}^2}$.

Now we are in the position to apply the above method to some concrete 
physical systems:
As a first example, we consider atomic Bose-Einstein condensates.
In the dilute-gas limit, the quantum phase and density fluctuations 
are small and can be treated as linear perturbations.
For wavelengths far above the healing length, the effective action 
of the phase fluctuations $\Phi$ reads (${\hbar=1}$)
\bea
\label{bec}
{\cal L}_{\rm eff}=\frac{1}{2}
\left(\frac{1}{g}\,\dot\Phi^2-\frac{\varrho_0}{m}\,(\na\Phi)^2\right)
\,,
\ea
where $\varrho_0$ is the background density of the condensate, 
$m$ the mass of the atoms, and $g$ the time-dependent coupling strength  
representing the inter-particle repulsion $g>0$ or attraction $g<0$.
Obviously, a homogeneous condensate becomes unstable for attractive 
interactions (i.e., $g_c=0$) and  this scenario ${g\downarrow0}$ 
corresponds to case~{\bf A}, i.e., an expanding universe.
Hence sweeping through the phase transition at $g_c=0$ by
means of a time-dependent external magnetic field (Feshbach resonance)
with a finite velocity generates a $k^{-4/3}$-spectrum of the
two-point phase-phase correlation function. 
However, it might be difficult to measure the phase after the 
phase transition (collapse of condensate due to molecule formation), 
whereas the frozen density fluctuations with large wavelengths should  
easily be measurable. 
The density fluctuations are just the canonically conjugated momentum 
field $\Pi$ and can be calculated analogously using the duality in 
Eqs.~(\ref{eom-phi}) and (\ref{eom-pi}), but with an additional factor 
of $k$ in Eq.~(\ref{hankel}).
Hence the spectrum of the density-density correlation function behaves 
as $k^{+4/3}$.
Note that this behavior is consistent with the amplification/suppression
of quantum fluctuations by squeezing which maintains the minimal
Heisenberg uncertainty of the ground state, i.e., 
${\Delta q_k\Delta p_k=\hbar/2}$.  

As a second example, let us study a simple 1+1 dimensional model of the 
electromagnetic field coupled to a linear medium via the magnetic component 
\bea
\label{em}
{\cal L}_{\rm eff}
=
\frac{1}{2}\left(E^2-B^2+\dot\Psi^2-\Omega^2\Psi^2+gB\Psi\right)
\,,
\ea
with the electromagnetic field being governed by the potential $A$ 
via ${E=\partial_t A}$ and ${B=\partial_x A}$.
The field $\Psi$ describes the (linearized and localized) dynamics of 
the medium with the plasma frequency $\Omega$ and $g$ denotes the coupling 
(magnetic dipole moment).
Averaging over the degrees of freedom $\Psi$ of the medium, the 
low-energy effective theory for the macroscopic electromagnetic field
yields the permeability $1/\mu=1-g^2/\Omega^2$
(which corresponds to inserting the adiabatic 
solution ${\Psi\approx gB/\Omega^2}$ back into the action).
Hence there is a phase transition at the critical value of 
the coupling $g_c=\Omega$ after which the medium develops a spontaneous
magnetization and the linearized description above breaks down.
Identifying $A=\Phi$, this scenario corresponds to case~{\bf B}.
Hence the frozen two-point spectra behave 
as $k^{-2/3}$ for $A$, 
and thus $k^{4/3}$ for $B$ (and $\Psi$), 
and finally $k^{2/3}$ for $E=\Pi$
(again respecting ${\Delta q_k\Delta p_k=\hbar/2}$)
if we sweep through the critical point with a finite velocity.
Note that, for the source-free macroscopic electromagnetic field, one
can also introduce the dual potential $\Lambda$ 
via ${H=B/\mu=\partial_t\Lambda}$ and ${D=E=\partial_x\Lambda}$, 
which explicitly incorporates the duality in Eqs.~(\ref{eom-phi}) and 
(\ref{eom-pi}), i.e., case~{\bf B} $\to$ case~{\bf A}, 
and consequently has a $k^{-4/3}$-spectrum.

Finally, as a third example, we consider a very simple version of the 
Heisenberg model
\bea
\label{Heisenberg}
H=-g(t)\sum\limits_{<ij>}\f{\sigma}_i\cdot\f{\sigma}_j
\,,
\ea
with $\f{\sigma}_i$ being the Pauli spin-1/2 operators for each  
lattice site $i$ and $<ij>$ denoting the sum over nearest and 
next-to-nearest neighbors, for example. 
The ferromagnetic
state ${\ket{\Psi_<}=\ket{\dots\uparrow\uparrow\uparrow\uparrow\dots}}$ 
is the ground state for $g>0$ and breaks the $O(3)$ invariance of the 
Hamiltonian; thus the spin-waves (magnons) are gap-less 
(Goldstone theorem).
Obviously $g=0$ is the critical point here, after which the energy of 
the magnons becomes negative.
In view of the global factor $g(t)$, this scenario is an example for 
case~{\bf C} with ${\alpha=\beta=g}$, i.e., ${a=b}$.
The observation that the spectrum of the fluctuations is not disturbed 
by the dynamics of $g(t)$ can also be explained by the fact that the 
ferromagnetic state ${\ket{\dots\uparrow\uparrow\uparrow\uparrow\dots}}$
is an exact eigenstate of the Hamiltonian, i.e., even the decay 
from $\ket{\Psi_<}$ to $\ket{\Psi_>}$ is impossible with the exact 
Hamiltonian in Eq.~(\ref{Heisenberg}) and requires some disturbances. 

In summary, the analogy between the excitations $\Phi$ and quantum fields 
in an expanding/contracting universe allows us to apply universal geometrical 
concepts such as horizons to a rather general class of quantum systems 
approaching the critical point.
Near the phase transition, adiabaticity breaks down 
(since the energy gap vanishes) and the system does not stay in its 
(instantaneous/adiabatic) ground state in general.
The spectrum of the two-point correlation function of the quantum 
fluctuations frozen out at the phase transition 
(which can be the seeds for pattern formation etc.) 
can be derived quite independent of the microscopic details of the
considered system and is basically determined by the characteristic
exponents $a$ and $b$ (universal behavior).
The strength of the frozen fluctuations, however, and their dynamics
after crossing the critical point (nonlinear instabilities etc.) 
depend on the explicit microscopic structure 
(e.g., the diluteness parameter in Bose-Einstein condensates).

As an outlook, one might compare phase transitions 
(such as the examples considered above) to ``real'' cosmic inflation which 
is part of the present standard model of cosmology.
Interestingly, phase transitions reproduce many features of 
inflation:
The decay of $\ket{\Psi_<}$ down to $\ket{\Psi_>}$ at $g>g_c$   
(breakdown of adiabaticity) releases energy (analogous to reheating). 
Phase transitions display universal behavior (no fine-tuning) 
in the sense that initial small-scale deviations from $\ket{\Psi_<}$ 
are not important after the transition $g>g_c$.
Similarly, the large-scale homogeneity may be explained naturally. 
However, even though quantum fluctuations generate small inhomogeneities
in both scenarios (phase transitions and ``real'' cosmic inflation),
none of the examples considered above reproduces the correct 
scale-invariant $1/k^3$-spectrum.
However, that is not surprising as the examples considered above break 
many symmetries we observe in the real universe, e.g., they possess a 
preferred frame and do not respect the principle of equivalence etc. 
If we {\em demand} that the effective action (at least at low energies)
does not single out a locally preferred frame 
(e.g., the two-point 
function ${\langle\hat\Phi(\ul x)\hat\Phi(\ul x')\rangle}$ 
depends on the Ricci scalar etc.)
there are only two possibilities:
firstly, the effectively flat space-time with ${a=b}$ leading to 
an undisturbed $1/k$-spectrum, and, secondly, the remaining nontrivial 
combination ${a+3b=-4}$ exactly corresponds to the de Sitter metric 
and thus reproduces the scale-invariant $1/k^3$-spectrum, 
cf.~Eq.~(\ref{nu}) and~\cite{2compBEC}. 
The second version allows for more additional symmetries:
We may assume a constant velocity of propagation for the $\Phi$-mode
\bea
{\cal A}=\int dt\,d^3r\,\frac{1}{2}\,
\frac{\dot\Phi^2-(\na\Phi)^2}{t^2}
\,,
\ea
and the resulting action turns out to be scale-invariant: 
${{\cal A}[t\to\lambda t,\f{r}\to\lambda\f{r}]={\cal A}[t,\f{r}]}$.
(There are also other motivations for the above form of the action
such as the principle of equivalence, which shall not be discussed 
here.)
One would expect the dynamics of this action, which has been 
motivated by {\em demanding} the above symmetries, to be generated by 
the back-reaction, which has been omitted so far and which respects 
these symmetries.
These interesting findings entice the question/speculation of whether
cosmic inflation itself might perhaps have been such a phase transition.

\acknowledgments

The author acknowledges valuable conversations with 
K.~Becker, U.~R.~Fischer, P.~Stamp, M.~Uhlmann, W.~G.~Unruh, and Y.~Xu.
This work was supported by the Emmy-Noether Programme of the 
German Research Foundation (DFG) under grant No.~SCHU 1557/1-1.  
Further support by the COSLAB programme of the ESF, 
the Pacific Institute of Theoretical Physics, 
and the programme EU-IHP ULTI is also gratefully acknowledged. 


\end{document}